\documentclass[preprint,showpacs,preprintnumbers,amsmath,amssymb]{revtex4}

\usepackage{graphicx}

\begin{document}

\title{
Negative time delay of light by a gravitational concave lens 
}
\author{Koki Nakajima}
\author{Koji Izumi}
\author{Hideki Asada} 
\affiliation{
Faculty of Science and Technology, Hirosaki University,
Hirosaki 036-8561, Japan} 
\date{\today}

\begin{abstract} 
Gravitational lens models, 
some of which might act as if a concave lens, 
have been recently investigated by using a static and spherically 
symmetric modified spacetime metric that 
depends on the inverse distance to the $n$-th power 
[Kitamura, Nakajima and Asada, PRD 87, 027501 (2013)]. 
We reexamine the time delay of light 
in a gravitational concave lens as well as a gravitational convex one. 
The frequency shift due to the time delay is also investigated. 
We show that the sign of the time delay in the lens models 
is the same as that of the deflection angle of light. 
The size of the time delay decreases with increase in 
the parameter $n$. 
We discuss also possible parameter ranges that are relevant to 
pulsar timing measurements in our galaxy. 
\end{abstract}

\pacs{04.40.-b, 95.30.Sf, 98.62.Sb}

\maketitle

\section{Introduction}
Gravitational effects on the light propagation provide 
powerful tools for investigating extrasolar planets, dark matter 
and dark energy in modern astronomy and cosmology. 
The effects are important also in theoretical physics 
particularly for studying a null structure of a spacetime. 
A rigorous form of the bending angle has been studied 
in understanding a strong gravitational field 
\cite{Frittelli, VE2000, Virbhadra, VNC, VE2002, VK2008, ERT, Perlick}. 
For example, 
strong gravitational lensing in a Schwarzschild black hole 
was considered by Frittelli, Kling and Newman \cite{Frittelli},  
by Virbhadra and Ellis \cite{VE2000} 
and more comprehensively by Virbhadra \cite{Virbhadra}; 
Virbhadra, Narasimha and Chitre \cite{VNC}
studied distinctive lensing features of naked singularities. 
Virbhadra and Ellis \cite{VE2002} 
and Virbhadra and Keeton \cite{VK2008} 
later described 
the strong gravitational lensing by naked singularities; 
Eiroa, Romero and Torres \cite{ERT} treated 
Reissner-Nordstr\"om black hole lensing; 
Perlick \cite{Perlick} discussed the lensing 
by a Barriola-Vilenkin monopole 
and also that by an Ellis wormhole \cite{Ellis,Morris1,Morris2}, 
the latter of which had been discussed long before 
(for instance, \cite{CC, Clement}). 
Moreover, it has been investigated as an observational probe 
of violations of some energy conditions 
\cite{Safonova, Visser, Shatskii, Perlick, Nandi, Abe, Toki, Tsukamoto, 
Tsukamoto2, Yoo}. 
There exist several forms of the deflection angle by the Ellis wormhole 
\cite{Perlick, Nandi, DS, BP, Abe, Toki, Tsukamoto}. 
A reason for such differences has been clarified by 
several authors \cite{Nakajima, Gibbons}. 
Furthermore, Tsukamoto and his collaborators \cite{Tsukamoto2014} 
have investigated the lensing in 
Tangherlini spacetime as an extension of Schwarzschild one 
to arbitrary dimensions; 
Horvath, Gergely, and Hobill \cite{HGH} 
studied lensing effects with negative convergence 
by so-called tidal charges in the Dadhich et al. solution 
\cite{Dadhich}, 
where, for a brane world black hole, the tidal charge 
arises from tidal forces acting at the brane-bulk boundary. 
Note that negative convergence in this case does not require 
any exotic matter. 
It comes from the Weyl curvature in higher dimensions. 
In addition, Gibbons and Kodama \cite{GK} have shown 
that curvature-regular asymptotically flat solitons with negative mass 
are contained in the Myers-Perry family, 
although the soliton solutions in the odd spacetime dimensions 
might not correspond to astrophysical objects. 
Kunz and her collaborators have constructed 
traversable wormholes 
in the dilatonic Einstein-Gauss-Bonnet theory \cite{Kunz2011, Kunz2012}, 
in which they do not require any form of exotic matter for their existence. 
Therefore, it is interesting to study gravitational lenses 
for probing an exotic energy or new physics. 

Furthermore, the arrival time delay of light 
in Schwarzschild spacetime, 
which is called Shapiro time delay, 
is another gravitational effect on light in a curved spacetime. 
The time delay effect has successfully tested 
the Einstein's theory \cite{Will}. 
A significant improvement was done by using Doppler tracking 
of the Cassini spacecraft on its way to the Saturn \cite{Cassini}. 
Gravitational time delay in some modified gravity models 
has been discussed by assuming a small perturbation 
around the Schwarzschild spacetime 
in the weak field approximation \cite{Asada2008}; 
Virbhadra and Keeton \cite{VK2008} and 
DeAndrea and Alexander \cite{DA} have discussed time delay 
by naked singularities to test the cosmic censorship hypothesis. 
Time delay by strongly naked singularities are negative, 
where strongly naked singularities are those that 
are not covered by any photon sphere. 

\begin{figure}
\includegraphics[width=10cm]{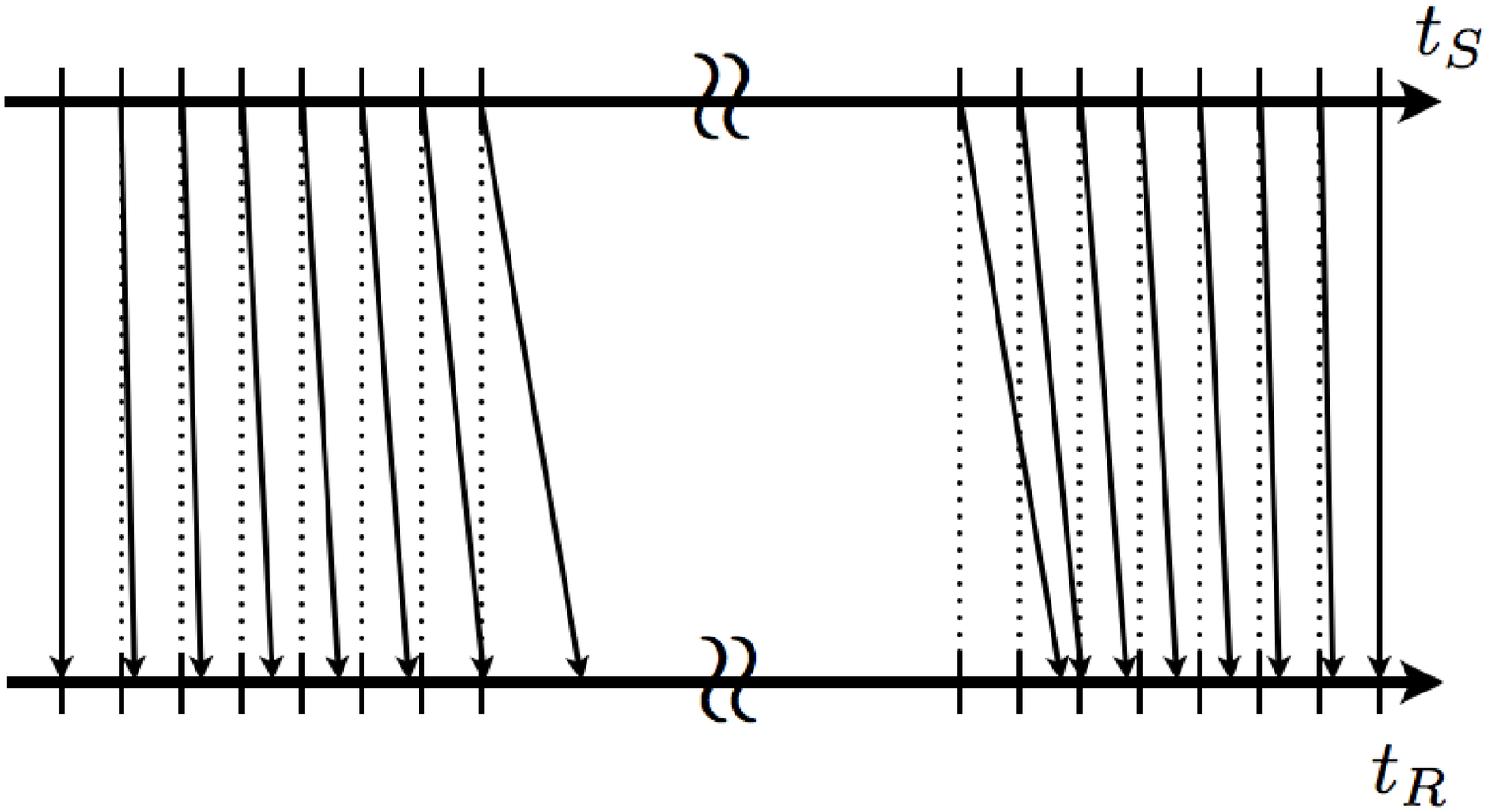}
\caption{
Frequency shift due to the gravitational time delay. 
There are two horizontal axes: 
One is corresponding to $t_S$ that is the time measured 
by a clock at the source of light, while 
the other is indicating $t_R$ that is the time measured 
by the receiver. 
The dotted vertical lines and solid ones are corresponding 
to the no-delay case and the time delay case, respectively. 
Signals of light are emitted with the same time interval, 
so that the intrinsic frequency of the signal can be constant. 
If the source is approaching to the lens object, 
the gravitational time delay of each signal 
is increasing with time and hence the observed frequency 
is lower than the intrinsic one, 
If the source is receding from the lens object, 
on the other hand, the delay is decreasing and hence 
the observed frequency is higher than the intrinsic one. 
}
\label{figure-1}
\end{figure}

Therefore, it will be interesting to study 
the time delay of light for understanding 
of other exotic lens models. 
Furthermore, the frequency shift is caused by a change 
of the time delay with time, 
if the emitter of light moves. 
Note that this frequency shift is different from 
the gravitational redshift due to a difference between clocks 
at two places. 
See Figure \ref{figure-1}, in which the frequency shift 
is caused by the time delay of each signal. 
Actually, it is the frequency shift by the gravitational time delay 
that was measured by the Cassini spacecraft experiment \cite{Cassini}. 
Hence, the present paper will study both the arrival time delay and 
the induced frequency shift. 
In particular, 
we reexamine the spacetime model that has been discussed 
by Kitamura et al. \cite{Kitamura} to study 
exotic gravitational lensing effects. 
This spacetime metric depends on the inverse distance 
to the power of positive $n$ 
in the weak field approximation. 
The Schwarzschild spacetime and the Ellis wormhole correspond to 
$n=1$ and $n=2$, respectively, so that these spacetimes 
can be expressed as a one-parameter family. 
This one-parameter model expresses a spherical mass distribution 
that may be related with violations of some energy conditions 
for a certain parameter region. 
Note that Birkhoff's theorem could say that cases $n \neq 1$ 
might be non-vacuum,  
if the models were interpreted in the framework of 
the standard Einstein equation. 
This lens model has suggested possible demagnification of lensed images 
\cite{Kitamura}, 
radial shear \cite{Izumi}, 
and an anomaly in centroid motions of images \cite{Kitamura2014}. 

This paper is organized as follows. 
Section II briefly summarizes the basics 
of the inverse-power lens models \cite{Kitamura,Izumi}. 
In Section III, we discuss time delay and frequency shift 
in the modified spacetime.  
Possible parameter ranges relevant to pulsar timings 
also are studied. 
Section IV is devoted to Conclusion.


\section{Modified spacetime model with an inverse power law} 
\subsection{Modified metric and the bending angle of light}
The present paper follows Kitamura et al. \cite{Kitamura}, 
which assumes that 
an asymptotically flat, static and spherically symmetric 
modified spacetime could depend on 
the inverse distance to the power of positive $n$ 
in the weak field approximation. 
We consider the light propagation through a four-dimensional spacetime, 
though the whole spacetime may be higher dimensional. 
The four-dimensional spacetime metric is \cite{Kitamura} 
\begin{equation}
ds^2=-\left(1-\frac{\varepsilon_1}{r^n}\right) c^2 dt^2
+\left(1+\frac{\varepsilon_2}{r^n}\right)dr^2
+r^2(d\Theta^2+\sin^2\Theta d\phi^2) 
+O(\varepsilon_1^2, \varepsilon_2^2, \varepsilon_1 \varepsilon_2) ,  
\label{ds}
\end{equation}
where $r$ is the circumference radius and 
$\varepsilon_1$ and $\varepsilon_2$ are small book-keeping 
parameters in iterative calculations. 
This model under study might describe 
a regular distribution of matter for $n \neq 1$ 
(See \cite{Izumi,Kitamura2014} on this point). 
The weak-field approximation means  
$\varepsilon_1 / r^n \ll 1$ 
and 
$\varepsilon_2 / r^n \ll 1$. 
Namely, we study a far field from the lens object as 
$r \gg \varepsilon_1^{1/n}$ and $r \gg \varepsilon_2^{1/n}$. 
Note that Eq. (\ref{ds}) is not valid 
in the strong field near $r=0$ 
(Please see a note in \cite{Kitamura2014} for more detail). 
Here, $\varepsilon_1$ and $\varepsilon_2$ 
may be either positive or negative, respectively. 
Negative $\varepsilon_1$ and $\varepsilon_2$ for $n=1$ 
correspond to a negative mass (in the linearized Schwarzschild metric). 

Without loss of generality, we focus on 
the equatorial plane $\Theta = \pi/2$, 
since the spacetime is spherically symmetric. 
The deflection angle of light 
becomes at the linear order \cite{Kitamura} 
\begin{align}
\alpha
&=\dfrac{\varepsilon}{b^n}\int_0^{\frac{\pi}{2}} \cos^n\Psi d\Psi 
+O(\varepsilon^2) , 
\label{alpha}
\end{align}
where the integral is positive definite, 
$b$ denotes the impact parameter of the light ray, 
we denote $\varepsilon \equiv n \varepsilon_1 + \varepsilon_2$, 
and we define $\Psi$ by $r_0/r = \cos\Psi$ for the closest approach $r_0$. 
By absorbing the positive integral 
into the parameter $\varepsilon$, we rewrite the linear-order 
deflection angle simply as 
$\alpha = \bar\varepsilon/b^n$, 
where the sign of $\bar\varepsilon$ is the same as that of $\varepsilon$. 
This deflection angle recovers 
the Schwarzschild ($n=1$) and Ellis wormhole ($n=2$) cases. 
$\varepsilon > 0$ and $\varepsilon < 0$ correspond to 
positive deflection of light and negative one, respectively. 
Hence, let us call them a gravitational convex lens and 
a gravitational concave one, respectively. 
It is unlikely that there exists any photon sphere 
in the concave model. 
For instance, see Eq. (7) in Ref. \cite{Kitamura2014}, 
which suggests $\varepsilon_1 > 0$ 
for the existence of a photon sphere. 

We mention an effective mass. 
A simple application of the standard lens theory \cite{SEF} 
suggests that the deflection angle of light in the form of 
$\alpha = \bar\varepsilon/b^n$ 
corresponds to a convergence (scaled surface mass density) as 
\begin{equation}
\kappa(b) = \frac{\bar\varepsilon (1-n)}{2} \frac{1}{b^{n+1}} , 
\label{kappa}
\end{equation}
which implies an extended spherical distribution of matter (or energy)  
for $n \neq 1$ and a singular source only for $n = 1$.

For the weak-field Schwarzschild case ($n = 1$), 
it follows that the convergence vanishes. 
If $\varepsilon > 0$ and $n>1$, 
or if $\varepsilon < 0$ and $n<1$, 
the effective surface mass density of the lens object 
is interpreted as negative in the framework 
of the standard lens theory \cite{Kitamura}. 
For these two cases, the matter (and energy) need to be exotic. 
If $\varepsilon < 0$ and $n>1$, 
or if $\varepsilon > 0$ and $n<1$, 
on the other hand, 
the convergence is positive almost everywhere 
except for the central singularity 
and hence exotic matter (and energy) are not required in the framework 
of the standard lens theory, in spite of the gravitational repulsion on 
light rays. 
Note that   
convex lenses ($\varepsilon > 0$) and concave ones ($\varepsilon < 0$) 
in the above two-parameter models do not have a one-to-one correspondence 
to positive convergence $\kappa > 0$ and negative one $\kappa < 0$.

\section{Time delay and frequency shift}
\subsection{Time delay of a light signal} 
\begin{figure}
\includegraphics[width=12cm]{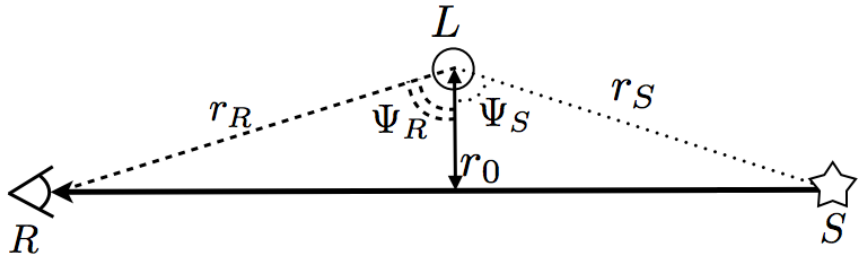}
\caption{
Schematic figure for a configuration of the source (emitter) 
of a signal of light, the receiver of the signal, and the lens. 
They are denoted by $S$, $R$ and $L$, respectively. 
The angles corresponding to the source and the receiver are 
denoted by $\Psi_S$ and $\Psi_R$. 
}
\label{figure-2}
\end{figure}

By using $ds^2 = 0$ for a light signal, 
we obtain the orbit equation as 
\begin{equation}
\left(\frac{dr}{dt}\right)^2
=\left(1-\dfrac{\varepsilon_1}{r^n}\right)
\left(1-\dfrac{\varepsilon_2}{r^n}\right)
\left[1-\dfrac{b^2}{r^2}\left(1-\dfrac{\varepsilon_1}{r^n}\right)\right] 
+O(\varepsilon_1^2, \varepsilon_2^2, \varepsilon_1 \varepsilon_2) ,
\label{orbit}
\end{equation}
where the impact parameter $b$ is related with 
the closest approach $r_0$ as 
\begin{equation}
b^2=\cfrac{r_0^2}{1-\cfrac{\varepsilon_1}{r_0^n}}.
\label{b}
\end{equation}
The time-of-flight of a photon from the source (denoted by $S$) 
to the receiver (denoted by $R$) becomes 
\begin{eqnarray}
t(S \to R)
&\equiv& \int_{r_S}^{r_R} dt 
\nonumber\\
&=&\int_{r_S}^{r_R} 
\left(1-\dfrac{r_0^2}{r^2}\right)^{-\frac{1}{2}}
\left(
1-\dfrac{\dfrac{r_0^2}{r^2}\dfrac{\varepsilon_1}{r_0^n}
\left(1-\dfrac{r_0^n}{r^n}
\right)}
{1-\dfrac{r_0^2}{r^2}}
-\dfrac{\tilde{\varepsilon}}{r^n}
\right)^{-\frac{1}{2}}dr
+O(\varepsilon_1^2, \varepsilon_2^2, \varepsilon_1 \varepsilon_2) ,
\label{traveltime}
\end{eqnarray}
where we define $\tilde{\varepsilon}\equiv\varepsilon_1+\varepsilon_2$.  
See Figure \ref{figure-2} for the photon path. 

Subtracting the time in the flat spacetime from it 
provides the time delay at the linear order as 
\begin{equation}
\delta t = 
\frac{1}{r_0^{n-1}}
\int_{\Psi_{S}}^{\Psi_{R}} 
\left(\dfrac{\varepsilon_1(1-\cos^n\Psi)}{\sin^2\Psi}
+\tilde{\varepsilon}\cos^{n-2}\Psi\right)d\Psi , 
\label{delay}
\end{equation}
where $\Psi_R$ and $\Psi_S$ correspond to the direction from the lens 
to the receiver and that to the source of light, respectively.

For general $n$, the integral in Eq. (\ref{delay}) is not always 
expressed in terms of elementary functions. 
Hence, numerical computations are required for Eq. (\ref{delay}). 
If $n$ is an integer, however, the integration of powers 
of trigonometric functions can be done \cite{GR}. 
For $n=1$, Eq. (\ref{delay}) becomes 
\begin{equation}
\delta t_1
=\varepsilon_1\left[
\dfrac{r_R}{x_R}+\dfrac{r_S}{x_S}
-\left(\dfrac{r_0}{x_R}+\dfrac{r_0}{x_S}\right)
\right]
+\tilde{\varepsilon}\left[
2\ln\dfrac{(r_R+x_R)(r_S+x_S)}{r_0^2}\right], 
\label{delay-n1}
\end{equation}
which agrees with the Shapiro delay formula 
if $\varepsilon_1 = \varepsilon_2$ equals to Schwarzschild radius. 
Here, we define $x_R \equiv \sqrt{r_R^2 - r_0^2}$ and 
$x_S \equiv \sqrt{r_S^2 - r_0^2}$. 

For $n=2$, we obtain 
\begin{equation}
\delta t_2
=\dfrac{\varepsilon_1+\tilde{\varepsilon}}{r_0}
\left(\arccos\dfrac{r_0}{r_R}+\arccos\dfrac{r_0}{r_S}\right).
\label{delay-n2}
\end{equation}

Next, we consider the case of an even integer ($n=2p$), 
where $p$ is a positive integer. 
Then, we obtain 
\begin{eqnarray}
\delta t_{2p}
&=&
\dfrac{\varepsilon_1}{c r_0^{2p-1}}
\left\{
    - \cot \Psi_R
    + \dfrac{\cos^{2p+1} \Psi_R}{\sin \Psi_R}
    + \dfrac{(2p-1)!!}{(2p-2)!!} \sin \Psi_R
    \sum^{P-1}_{r=0} \dfrac{(2p-2r-2)!!}{(2p-2r-1)!!} \cos^{2p-2r-1} \Psi_R
\right.
\nonumber
\\
&\ &
\hspace{13mm}
\left.
    + \cot \Psi_S
    - \dfrac{\cos^{2p+1} \Psi_S}{\sin \Psi_S}
    - \dfrac{(2p-1)!!}{(2p-2)!!} \sin \Psi_S
    \sum^{P-1}_{r=0} \dfrac{(2p-2r-2)!!}{(2p-2r-1)!!} \cos^{2p-2r-1} \Psi_S
\right\}
\nonumber
\\
&\ &
+
\dfrac{\tilde{\varepsilon}}{c r_0^{2p-1}}
\left\{
      \dfrac{(2p-3)!!}{(2p-2)!!} \sin \Psi_R
      \sum^{P-2}_{r=0} \dfrac{(2p-2r-4)!!}{(2p-2r-3)!!} \cos^{2p-2r-3} \Psi_R
      + \dfrac{(2p-3)!!}{(2p-2)!!} \Psi_R
\right.
\nonumber
\\
&\ &
\hspace{13mm}
\left.
      - \dfrac{(2p-3)!!}{(2p-2)!!} \sin \Psi_S
      \sum^{P-2}_{r=0} \dfrac{(2p-2r-4)!!}{(2p-2r-3)!!} \cos^{2p-2r-3} \Psi_S
      - \dfrac{(2p-3)!!}{(2p-2)!!} \Psi_S
\right\} , 
\label{delay-neven}
\end{eqnarray}
where the subscript $2p$ indicates the case of $n=2p$ and 
$(2p-1)!!$ means $(2p-1)(2p-3) \cdots 1$. 

Next, let us consider the case of $n=2p+1$. 
Then, we obtain 
\begin{eqnarray}
\delta t_{2p+1}
&=&
\dfrac{\varepsilon_1}{c r_0^{2p}}
\left\{
    - \cot \Psi_R
    + \dfrac{\cos^{2p+2} \Psi_R}{\sin \Psi_R}
    + \dfrac{(2p)!!}{(2p-1)!!} \sin \Psi_R
      \sum^P_{r=0} \dfrac{(2p-2r-1)!!}{(2p-2r)!!} \cos^{2p-2r} \Psi_R
\right.
\nonumber
\\
&\ &
\hspace{10mm}
\left.
    + \cot \Psi_S
    - \dfrac{\cos^{2p+2} \Psi_S}{\sin \Psi_S}
    - \dfrac{(2p)!!}{(2p-1)!!} \sin \Psi_S
      \sum^P_{r=0} \dfrac{(2p-2r-1)!!}{(2p-2r)!!} \cos^{2p-2r} \Psi_S
\right\}
\nonumber
\\
&\ &
+
\dfrac{\tilde{\varepsilon}}{c r_0^{2p}}
\left\{
      \dfrac{(2p-2)!!}{(2p-1)!!} \sin \Psi_R
      \sum^{p-1}_{r=0} \dfrac{(2p-2r-3)!!}{(2p-2r-2)!!} \cos^{2p-2r-2} \Psi_R
\right.
\nonumber
\\
&\ &
\hspace{10mm}
-
\left.
      \dfrac{(2p-2)!!}{(2p-1)!!} \sin \Psi_S
      \sum^{p-1}_{r=0} \dfrac{(2p-2r-3)!!}{(2p-2r-1)!!} \cos^{2p-2r} \Psi_S
\right\} . 
\label{delay-nodd}
\end{eqnarray}

Up to this point, integrations have been done without any approximation. 
For astronomical situations, 
the closest distance of light is much shorter than 
the distance from the lens to the source 
and that from the lens to the observer, 
so that we can take the limit as $r_S/r_0 \to \infty$ 
and $r_R/r_0 \to \infty$. 
This leads to $\Psi_R \to \pi/2$ and $\Psi_S \to - \pi/2$, 
so that the above expressions can be simplified. 
See also Figure \ref{figure-2}. 
For $n=2p$, we obtain 
\begin{eqnarray}
    \delta t_{2p} = \frac{\pi}{c} \dfrac{(2p-3)!!}{(2p-2)!!} 
\dfrac{2p \varepsilon_1 + \varepsilon_2}{r_0^{2p-1}} . 
\label{delay-neven-limit}
\end{eqnarray}
For $n=2p+1$ case, we obtain 
\begin{eqnarray}
    \delta t_{2p+1} = \frac{2}{c} \dfrac{(2p-2)!!}{(2p-1)!!} 
\dfrac{(2p + 1) \varepsilon_1 + \varepsilon_2}{r_0^{2p}} . 
\label{delay-nodd-limit}
\end{eqnarray}
Eqs. (\ref{delay-neven-limit}) and (\ref{delay-nodd-limit}) 
suggest that the time delay $\delta t$ is proportional to 
$n \varepsilon_1 + \varepsilon_2 = \varepsilon$ 
and also to $r_0^{-n}$, if $r_R$ and $r_S$ are large. 
Note that the sign of the time delay 
is the same as that of the deflection angle of light. 
The time delay $\delta t$ for $\varepsilon > 0$ 
is a downward-convex function of $r_0$ almost everywhere 
except for $r_0 = 0$, 
while $\delta t$ for $\varepsilon < 0$ is convex upward.

\begin{figure}
\includegraphics[width=12cm]{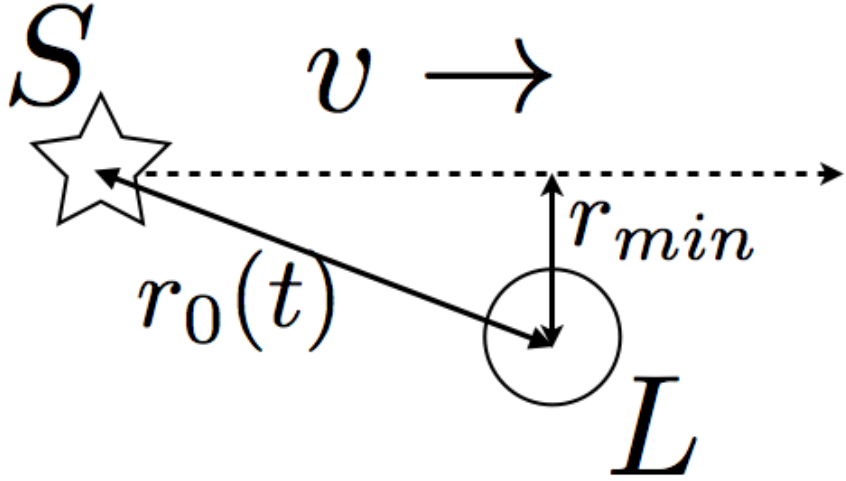}
\caption{
Schematic figure for a motion of the light emitter 
with respect to the lens, where their positions are projected 
onto the celestial sphere. 
The closest approach of light to the lens 
is a function of time denoted as $r_0(t)$, 
and 
its minimum is denoted as $r_{min}$. 
}
\label{figure-3}
\end{figure}

\begin{figure}
\includegraphics[width=10cm]{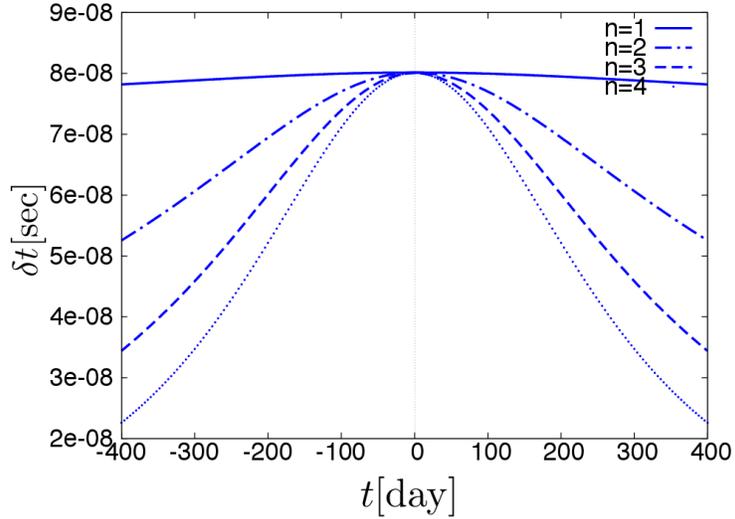}
\caption{
Time delay curves for $\varepsilon > 0$.
The solid, dot-dashed, dashed, and dotted curves 
correspond to $n=1$, $2$, $3$ and $4$, respectively. 
The horizontal axis denotes the time $t$ in days and 
the vertical axis means the time delay $\delta_t$ in seconds. 
Here, we assume $r_{min}$ is $40$ AU and $v = 200$ km/s. 
The lens is assumed to be a ten solar mass black hole 
for $n=1$ ($\varepsilon/r_{min} \sim 10^{-8}$), 
and the parameters for the other $n$ are chosen 
such that the peak height of the time delay curve can remain the same 
as each other. 
}
\label{figure-4}
\end{figure}

\begin{figure}
\includegraphics[width=10cm]{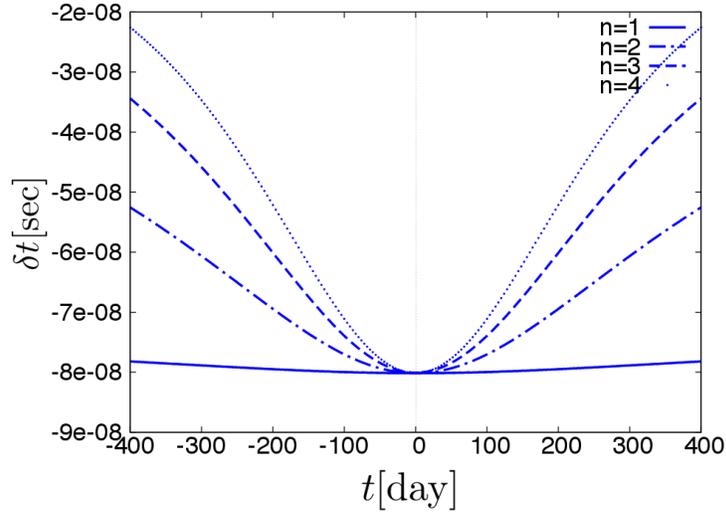}
\caption{
Time delay curves for $\varepsilon < 0$. 
Here, the parameter values are the same as those 
in Figure \ref{figure-4}, 
except for the sign of $\varepsilon$. 
}
\label{figure-5}
\end{figure}

Figure \ref{figure-3} shows schematically 
a motion of the signal emitter with respect to 
the lens object. 
We assume that the emitter and the lens object are in a linear motion 
during short-time observations, 
so that we can take $r_0(t) = \sqrt{r_{min}^2 + v^2 t^2}$.  
Here, $v$ denotes the transverse component of the 
relative velocity between the lens and the source 
with respect to the line of sight, 
$r_{min}$ denotes the minimum of $r_0$ and 
$t=0$ is chosen as the passage time of $r_0 = r_{min}$ 
without loss of generality. 
Figure \ref{figure-4} shows curves for positive time delays for 
$n=1, 2, 3$ and $4$ with $\varepsilon >0$.  
Figure \ref{figure-5} corresponds to negative delays 
for $\varepsilon <0$.

\subsection{Frequency shift}
Next, we consider the frequency shift caused by the time delay. 
In most astronomical situations, 
observers cannot know the exact time of the emission of light. 
Hence, the arrival time delay cannot be measured directly 
for a single image, 
although the arrival time difference between multiple images 
(e.g. distant quasars) are astronomical observables \cite{SEF}. 
Note that the round-trip time of a light signal 
is a direct observable for a spacecraft 
such as Voyager and Cassini. 
On the other hand, the frequency shift induced by the time delay 
becomes a direct observable, if the emitter of a light signal 
moves with respect to the lens object \cite{Cassini}. 

The frequency shift $y$ due to the time delay 
is defined as \cite{Cassini,Asada2008}
\begin{eqnarray}
y &\equiv& \frac{\nu(t) - \nu_0}{\nu_0} 
\nonumber\\
& =& - \dfrac{d(\delta t)}{dt} , 
\label{y}
\end{eqnarray} 
where $\nu_0$ denotes the intrinsic frequency of light 
and $\nu(t)$ means the observed one at time $t$. 
As Figure \ref{figure-1} suggests, 
$y < 0$ if the emitter of light approaches the lens 
and $y > 0$ if it recedes. 
In general, the expression for $y$ may become lengthy, 
though it is simplified for an integer $n$. 
First, we differentiate Eqs. (\ref{delay-neven}) 
and (\ref{delay-nodd}) with respect to time, 
where we use 
\begin {eqnarray}
    \dfrac{d \Psi_I}{dt} = 
- \dfrac{1}{\sqrt{r_I^2 - r_0^2}} \dfrac{dr_0}{dt} ,   
\end{eqnarray}
for $I = R, S$. 
Next, we take the limit as 
$\Psi_R \to \pi/2$ and $\Psi_S \to - \pi/2$. 

For $n=2p$, 
the frequency shift is obtained as 
\begin{eqnarray}
    y_{2p} = \frac{\pi}{c} \dfrac{(2p-1)!!}{(2p-2)!!} 
\dfrac{\varepsilon}{r_0^{2p+1}} v^2 t , 
\label{y2p}
\end{eqnarray}
and for $n=2p+1$, it becomes 
\begin{eqnarray}
    y_{2p+1} = \frac{2}{c} \dfrac{(2p)!!}{(2p-1)!!} 
\dfrac{\varepsilon}{r_0^{2p+2}} v^2 t .  
\label{y2p+1}
\end{eqnarray}

\begin{figure}
\includegraphics[width=10cm]{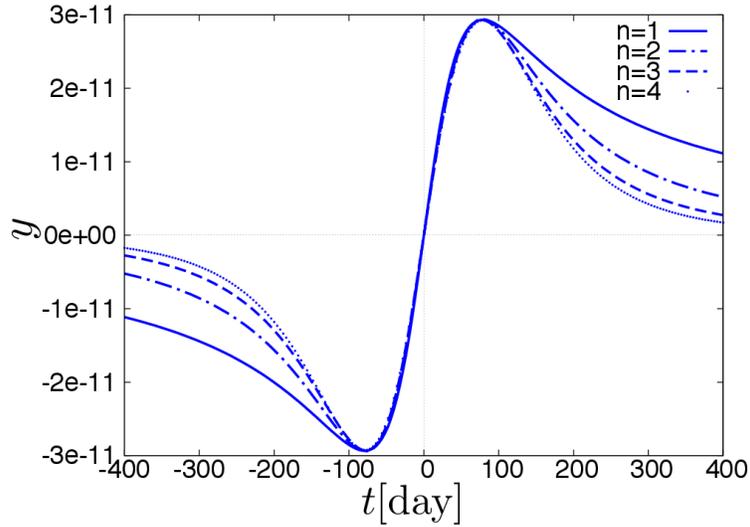}
\caption{
Frequency shift curves for $\varepsilon > 0$ 
corresponding to Figure \ref{figure-4}. 
Here, the parameter values for $n \neq 1$ are rearranged 
such that the peak height of the time delay curve can remain the same 
as each other. 
}
\label{figure-6}
\end{figure}

\begin{figure}
\includegraphics[width=10cm]{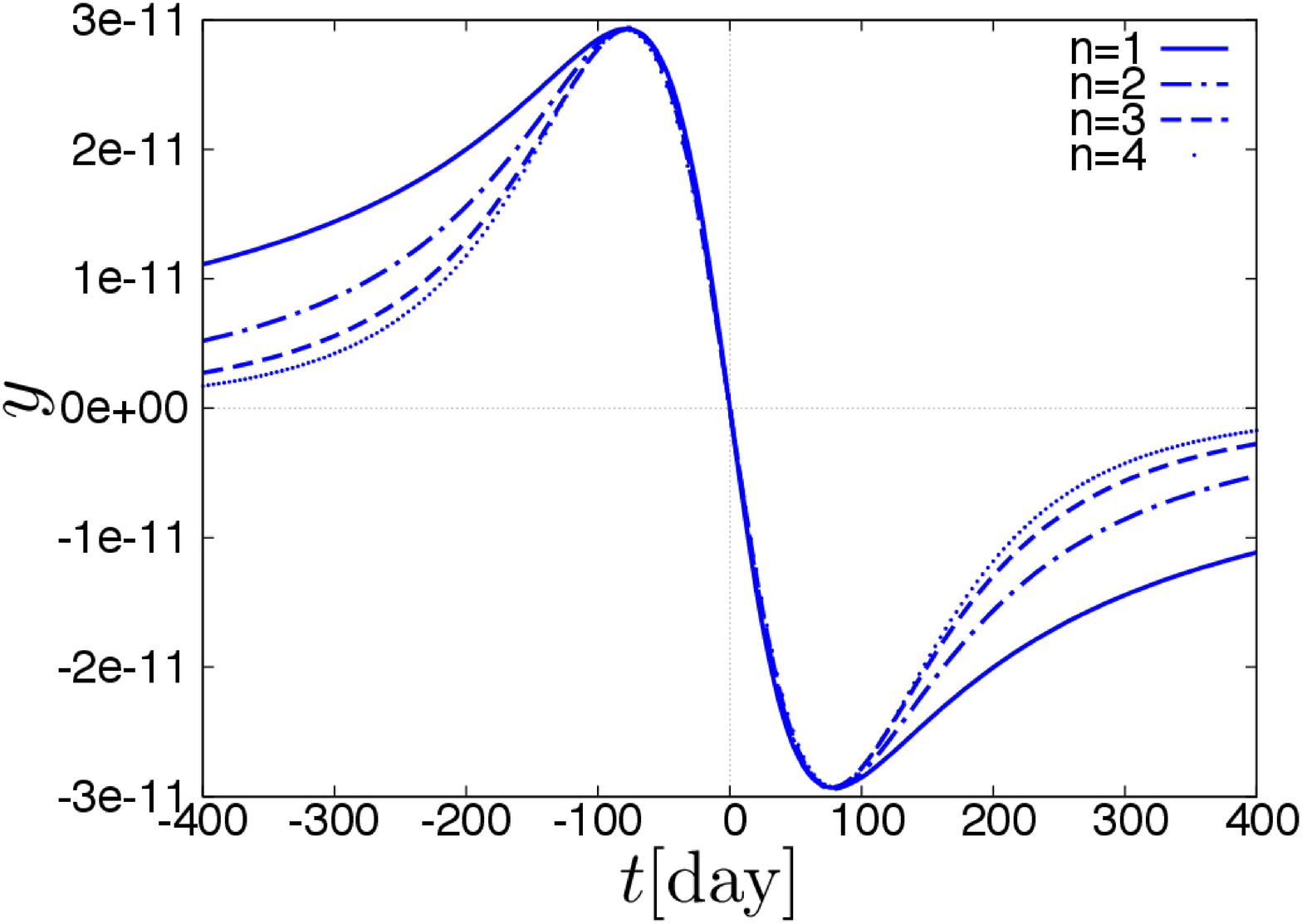}
\caption{
Frequency shift curves for $\varepsilon < 0$ 
corresponding to Figure \ref{figure-5}. 
Here, the parameter values are the same as those 
in Figure \ref{figure-6}, 
except for the sign of $\varepsilon$. 
}
\label{figure-7}
\end{figure}

Figure \ref{figure-6} for $\varepsilon > 0$ 
shows the frequency shift for $n=1, 2, 3$, and $4$. 
Here, the appropriate value of the parameter $r_0$ is chosen 
such that the peak location can remain the same with each other 
in order to enable us to see the behavior of $y$. 
The timing shift curve falls off more rapidly as the parameter $n$ 
increases. 
This is because the gravitational potential decays faster 
as $n$ becomes larger (See the spacetime metric in Eq. (\ref{ds})). 
Eqs. (\ref{y2p}) and (\ref{y2p+1}) suggest that 
$y \propto t/r_0^{n+1} \propto t^{-n}$ for large $|t|$. 
See Figure \ref{figure-7} for the $\varepsilon < 0$ case.

\subsection{Possible parameter ranges in pulsar timing method} 
There are two observables in time delay measurements. 
One is the amplitude of the time delay corresponding
to the curve height in Figures \ref{figure-4} and \ref{figure-6},
and the other is the duration corresponding to the curve width
in Figure \ref{figure-5} and \ref{figure-7}.
Eqs. (\ref{delay-neven-limit}) and (\ref{delay-nodd-limit}) suggest that
the size of the time delay is
\begin{equation}
\delta t \sim \frac{1}{c} \frac{\varepsilon}{r_0^n} .
\end{equation}
The duration of the time delay is roughly
\begin{equation}
T_{td} \sim \cfrac{\delta t}{\left( \cfrac{d(\delta t)}{dt}\right)} 
\sim \frac{r_0}{v} .
\end{equation}
Let us assume that $v$ is of the order of the typical rotational velocity
in our galaxy,
because there are no established theories 
on the motion of the exotic lens.
The current pulsar timing measurements are done a few times per year 
for each pulsar.
Th root-mean-squared residuals of the timing arrival dispersions are
nearly 100 - 1 nanoseconds (ns) that are depending on pulsars 
\cite{Pulsar,Yunes}.
The relevant value of the closest approach becomes
\begin{equation}
r_0 \sim 40 \mbox{AU}  
\left(\frac{T_{td}}{1 \mbox{year}}\right)
\left(\frac{v}{200 \mbox{km/s}}\right) ,
\end{equation}
and
the exotic lens potential is roughly
\begin{equation}
\frac{\varepsilon}{r_0^n} \sim 10^{-11}
 \left(\frac{\delta t}{100 \mbox{ns}}\right) .
\end{equation}
This is corresponding to the frequency shift $y \sim 10^{-15}$. 
This accuracy has been already achieved for a milli-second pulsar 
\cite{Pulsar,Yunes}. 

Furthermore, let us consider a possible constraint on the number
density of the exotic lens models.
One event would be detectable, 
if a lens object entered a volume of a cylinder as 
$V \sim \pi r_0^2 D_S$, where $D_S$ denotes the distance of the source
(pulsar) from
the observer.
The effective survey volume is proportional to both the number of the
observed pulsars (denoted as $N_p$) 
and the total duration of a pulsar timing project (denoted as $T_{pt}$).
Hence, the number density of the lens objects $\Omega_L$ 
would be constrained by no event detection as
\begin{equation}
\Omega_L < 10^3 \mbox{pc}^{-3} 
\left( \frac{40 \mbox{AU}}{r_0} \right)^2 
\left( \frac{1 \mbox{kpc}}{D_S} \right)
\left( \frac{10 \mbox{year}}{T_{pt}} \right) . 
\end{equation}
Although it seems very weak, this constraint 
might be interesting, because $n>1$ models are massless 
at the spatial infinity and thus it is unlikely that 
these exotic objects are constrained by other observations 
regarding stellar motions, galactic rotation and so on.

\section{Conclusion}
We examined the arrival time delay of light and 
the frequency shift in the lens model 
with an inverse power law. 
The time delay by a gravitational convex lens 
(i.e. positive deflection angle of light) 
would be positive,  
even if the lens model had negative convergence 
like Ellis wormholes. 
On the other hand, time delay by a gravitational concave lens 
might become negative, 
even if the convergence were positive. 

By using the Janis-Newman-Winicour metric, 
Virbhadra and Keeton \cite{VK2008} and 
DeAndrea and Alexander \cite{DA} have shown that 
negative time delay is caused in strongly naked singularities, 
where strongly naked singularities are those which are not 
covered within any photon spheres. 
The present model of a gravitational concave lens suggests that 
there is no photon spheres, though  
the present analysis is limited within the weak-field approximation. 
The previous works and the present one suggest that 
negative time delay might be related with 
whether there exists a photon sphere in a spacetime. 
This is left as a future work. 

Finally, we mention a comparison 
with the previous works \cite{Kitamura, Izumi, Kitamura2014}. 
They find distinctive features in lensing effects that 
are closely related with negative convergence. 
On the other hand, the present paper shows that negative 
time delay might be caused by a gravitational concave lens, 
even though the convergence is positive. 
Therefore, the present result might suggest that time delay 
in such a spacetime will bring us additional information 
independent of lensing effects. 
Further study along this course is left as a future work.

We would like to thank F. Abe, M. Bartelmann, T. Harada, 
S. Hayward, T. Kitamura, J. Kunz, K. Nakao, Y. Sendouda, R. Takahashi, 
N. Tsukamoto, and M. Visser 
for the useful conversations on the exotic lens models.

\newpage

\begin{thebibliography}{99}
\bibitem{Frittelli}
S. Frittelli, T. P. Kling, and E. T. Newman,
Phys. Rev. D {\bf 61}, 064021 (2000).
\bibitem{VE2000}
K. S. Virbhadra, and G. F. R. Ellis,
Phys. Rev. D {\bf 62}, 084003 (2000).
\bibitem{Virbhadra}
K. S. Virbhadra,
Phys. Rev. D {\bf 79}, 083004 (2009).
\bibitem{VNC}
K. S. Virbhadra, D. Narasimha, and  S. M. Chitre,
Astron. Astrophys. {\bf 337}, 1 (1998).
\bibitem{VE2002}
K. S. Virbhadra, and G. F. R. Ellis,
Phys. Rev. D {\bf 65}, 103004 (2002).
\bibitem{VK2008}
K. S. Virbhadra, and C. R. Keeton,
Phys. Rev. D {\bf 77}, 124014 (2008).
\bibitem{ERT}
E. F. Eiroa, G. E. Romero, and D. F. Torres,
Phys. Rev. D {\bf 66}, 024010 (2002).
\bibitem{Perlick}
V. Perlick,
Phys. Rev. D {\bf 69}, 064017 (2004).
\bibitem{Ellis}
H. G. Ellis, J. Math. Phys. (N.Y.) {\bf 14}, 104 (1973).
\bibitem{Morris1}
M. S. Morris, and K. S. Thorne,
Am. J. Phys. {\bf 56}, 395 (1988).
\bibitem{Morris2}
M. S. Morris, K. S. Thorne, and U. Yurtsever,
Phys. Rev. Lett. {\bf 61}, 1446 (1988).
\bibitem{CC}
L. Chetouani, and G. Cl\'ement,
Gen. Relativ. Gravit. {\bf 16}, 111 (1984).
\bibitem{Clement}
G. Cl\'ement,
Int. J. Theor. Phys. {\bf 23}, 335 (1984).
\bibitem{Visser}
M. Visser, {\it Lorentzian Wormholes: From Einstein to Hawking}
(AIP, New York, 1995).
\bibitem{Safonova}
M. Safonova, D. F. Torres, and G. E. Romero,
Phys. Rev. D {\bf 65}, 023001 (2001).
\bibitem{Shatskii}
A. A. Shatskii, Astron. Rep. {\bf 48}, 525 (2004).
\bibitem{Nandi}
K. K. Nandi, Y. Z. Zhang, and A. V. Zakharov,
Phys. Rev. D {\bf 74}, 024020 (2006).
\bibitem{Abe}
F. Abe, Astrophys. J. {\bf 725}, 787 (2010).
\bibitem{Toki}
Y. Toki, T. Kitamura, H. Asada, and F. Abe,
Astrophys. J. {\bf 740}, 121 (2011).
\bibitem{Tsukamoto}
N. Tsukamoto, T. Harada, K. Yajima,
Phys. Rev. D {\bf 86}, 104062 (2012).
\bibitem{Tsukamoto2}
N. Tsukamoto, and T. Harada,
Phys. Rev. D {\bf 87}, 024024 (2013).
\bibitem{Yoo}
C. M. Yoo, T. Harada, and N. Tsukamoto,
Phys. Rev. D {\bf 87}, 084045 (2013).
\bibitem{DS}
T. K. Dey, and S. Sen,
Mod. Phys. Lett. A, {\bf 23}, 953 (2008).
\bibitem{BP}
A. Bhattacharya, and A. A. Potapov,
Mod. Phys. Lett. A, {\bf 25}, 2399 (2010).
\bibitem{Nakajima}
K. Nakajima, and H. Asada,
Phys. Rev. D {\bf 85}, 107501 (2012).
\bibitem{Gibbons}
G. W. Gibbons, and M. Vyska,  Class. Quant. Grav. {\bf 29} 065016 (2012).
\bibitem{Tsukamoto2014}
N. Tsukamoto, T. Kitamura, K. Nakajima, and H. Asada,
arXiv:1402.6823 [gr-qc].
\bibitem{HGH}
Z. Horvath, L. Gergely, and D. Hobill,
Class. Quant. Grav. {\bf 27}, 235006 (2010).
\bibitem{Dadhich}
N. Dadhich, R. Maartens, P. Papadopoulos, and V. Rezania,
Phys. Lett. B, {\bf 487}, 1 (2000).
\bibitem{GK}
G. Gibbons, and H. Kodama,
Prog. Theor. Phys. {\bf 121}, 1361 (2009).
\bibitem{Kunz2011}
B. Kleihaus, J. Kunz, and E. Radu,
Phys. Rev. Lett. {\bf 106}, 151104 (2011).
\bibitem{Kunz2012}
P. Kanti, B. Kleihaus, and J. Kunz,
Phys. Rev D {\bf 85}, 044007 (2012).
\bibitem{Will}
C. M. Will, Living Rev. Relativity, {\bf 9}, 3 (2006),
[http://relativity.livingreviews.org/Articles/lrr-2006-3].
\bibitem{Cassini}B. Bertotti, L. Iess, and P. Tortora,
Nature, {\bf 425}, 374 (2003).
\bibitem{Asada2008}
H. Asada,
Phys. Lett. B, {\bf 661}, 78 (2008).
\bibitem{DA}
J. P. DeAndrea, and K. M. Alexander,
ArXiv:1402.5630 [gr-qc].
\bibitem{Kitamura}
T. Kitamura, K. Nakajima, and H. Asada,
Phys. Rev. D {\bf 87}, 027501 (2013).
\bibitem{Izumi}
K. Izumi, C. Hagiwara, K. Nakajima, T. Kitamura, and H. Asada,
Phys. Rev. D {\bf 88}, 024049 (2013).
\bibitem{Kitamura2014}
T. Kitamura, K. Izumi, K. Nakajima, C. Hagiwara, and H. Asada,
Phys. Rev. D {\bf 89}, 084020 (2014).
\bibitem{SEF}
P. Schneider, J. Ehlers, and E. E. Falco,
{\it Gravitational Lenses}
(Springer, New York, 1992).
\bibitem{GR}
I. S. Gradshteyn, and I. M. Ryzhik,
{\it Table of Integrals, Series and Products},
Seventh Edition, pages 151-160,
(Academic Press, New York, 2007).
\bibitem{Pulsar}
G. Hobbs, et al.,
Classical and Quantum Gravity, {\bf 27}, 8 (2010).
\bibitem{Yunes}
N. Yunes, and X. Siemens, 
Living Rev. Relativity, {\bf 16}, 9 (2006), \par
[http://relativity.livingreviews.org/Articles/lrr-2013-9].
\end{thebibliography}
\end{document}